# Thermal gravitational waves

C Sivaram and Kenath Arun[1]

Indian Institute of Astrophysics, Bangalore

**Abstract:** There is a lot of current interest in sources of gravitational waves and active ongoing projects to detect such radiation, such as the LIGO project. These are long wavelength, low frequency gravitational waves. LISA would be sensitive to much longer wavelengths and lower fluxes. However compact stellar objects can generate high frequency $(10^{16} - 10^{21} Hz)$ thermal gravitational radiation, which in the case of hot neutron stars can be high. Also white dwarfs and main-sequence stars can generate such radiation from plasma-Coulomb collisions. Again gamma ray bursts and relativistic jets could also be sources of such radiation. Terminal stages of evaporating black holes could also generate high frequency gravitational radiation. A comparative study is made of the thermal gravitational wave emission from all of the above sources, and the background flux is estimated. The earliest phases of the universe close to the Planck scale would also leave remnant thermal gravitational waves. The integrated thermal gravitational flux as the universe expands is also estimated and compared with that from all the discrete sources discussed above. Possible schemes to detect such sources of high frequency thermal gravitational radiation are discussed and the physical principles involved are elaborated.

---

[1] Christ Junior College, Bangalore

Due to Coulomb collisions in the core of the stars, thermal gravitational waves can be generated. These thermal gravitational waves can arise in white dwarfs and neutron stars due to the fermion collisions in the dense degenerate Fermi gas. Such high frequency thermal gravitational waves are also produced during the collisions in a gamma ray burst or in the jets of a rotating black hole.

### 1. Thermal gravitational waves from stellar cores

If $n_1, n_2$ are the number densities of gas particles undergoing collision with a $\frac{d\sigma_{12}}{d\Omega}$ differential scattering cross-section, with relative velocity $v_{12}$ and reduced mass $\mu_{12}$, then the power per unit volume per unit frequency interval is given by the quadrupole formula as: [1, 2]

$$\dot{E} = \left( \frac{32G}{5c^5} \mu_{12}^2 n_1 n_2 v_{12}^5 \sum \frac{d\sigma}{d\Omega} \sin^2 \theta \right) V \nu \qquad \ldots(1)$$

Where, $V$ is the volume of the stellar core and $\nu = \frac{kT_c}{h} \approx 10^{17} Hz$ is the frequency corresponding to the core temperature of the star of $T_c \approx 10^7 K$. The velocity of the particles at the core of temperature $T_c$ is given by:

$$v_{12} = \sqrt{\frac{3kT_c}{\mu_{12}}}.$$

For a core temperature of $T_c \approx 10^7 K$, the velocity of the particles is of the order of $\approx 8 \times 10^5 m/s$. For a star of density $\rho \approx 200 g/cc = 2 \times 10^5 kg/m^3$, the number density is given as: $n_1 = n_2 = \rho/m_P \approx 10^{32} m^{-3}$. The volume of the star is of the order of $V \approx 10^{27} m^3$.

And for a main sequence star [3], $\sum \frac{d\sigma}{d\Omega} \sin^2 \theta = \frac{e^4}{(8\pi\varepsilon_0)^2 \mu_{12}^2 v_{12}^4} \approx 5 \times 10^{-28} m^2 \qquad \ldots(2)$

Putting in these values we get the power of thermal gravitational waves emitted as, $\dot{E} \approx 10^9 Watt$ at a frequency of $\nu \approx 10^{17} Hz$.

The flux of thermal gravitational waves from the sun, received at earth is of the order of half a watt.

## 2. Thermal gravitational waves from compact stars

In the case of white dwarfs, the number density is of the order of $n_1 = n_2 \approx 10^{37} m^{-3}$, and the velocity corresponding to the white dwarf temperature of $T_c \approx 10^8 K$, is of the order of $\approx 2 \times 10^6 m/s$.

The volume of the white dwarf is of the order of $4 \times 10^{18} m^3$ and the frequency corresponding to the temperature $T_c \approx 10^8 K$ is $\nu = \dfrac{kT_c}{h} \approx 10^{18} Hz$.

And for a white dwarf, $\sum \dfrac{d\sigma}{d\Omega} \sin^2 \theta = \dfrac{e^4}{(8\pi\varepsilon_0)^2 \mu_{12}^2 v_{12}^4} \approx 10^{-29} m^2$ …(3a)

The power of thermal gravitational waves emitted by the white dwarf works out to be of the order of, $\dot{E} \approx 10^{12} Watt$ at the frequency of $\nu = 10^{18} Hz$

In the case of neutron stars, the number density is of the order of $n_1 = n_2 \approx 10^{44} m^{-3}$, and the velocity corresponding to the neutron star temperature of $T_c \approx 5 \times 10^{10} K$, is of the order of $\approx 5 \times 10^7 m/s$.

The volume of the neutron star is of the order of $4 \times 10^{12} m^3$ and the frequency corresponding to the temperature $T_c \approx 5 \times 10^{10} K$ is $\nu = \dfrac{kT_c}{h} \approx 10^{21} Hz$.

And for a neutron star, $\sum \dfrac{d\sigma}{d\Omega} \sin^2 \theta = \dfrac{e^4}{(8\pi\varepsilon_0)^2 \mu_{12}^2 v_{12}^4} \approx 10^{-33} m^2$ …(3b)

The power of thermal gravitational waves emitted by the neutron star works out to be of the order of, $\dot{E} \approx 10^{22} Watt$ at the frequency of $\nu = 10^{21} Hz$.

We assume collisions of neutrons described by hard sphere fermion model with scattering length of the order of $5 \times 10^{16} m$. We restrict to S-wave scattering since the de Broglie wavelength is large compared to this length. The integrated power density is given by:

$$P_g = \frac{8G}{5c^5}(3\pi^2 n)^{2/3} l^2 \left(\frac{M}{m_n}\right)\left(\frac{kT_c}{\hbar^{1/2}}\right)^4 \qquad \ldots(4)$$

Where, $M$ is the mass of the star, $m_n$ is the neutron mass, $n$ is the central number density and $T_c$ is the core temperature assumed much smaller than the Debye temperature $T_D$. Even for a newly formed hottest neutron star $T_c = 5 \times 10^{10} K < T_D (10^{14} K)$. [4]

## 3. Thermal gravitational waves from gamma ray bursts

Gamma-ray bursts (GRBs) are the most luminous physical phenomena in the universe known to the field of astronomy. They consist of flashes of gamma rays that last from seconds to hours, the longer ones being followed by several days of X-ray afterglow. [5,6]

Similar to the stars, in GRBs also, Coulomb collisions can result in the emission of high frequency gravitational waves.

The power of the thermal gravitational waves is given by the same expression as that for the stars, but the bulk properties will be altered by factors of $\Gamma$, due to the relativistic velocities encountered in GRBs.

$$\dot{E} = \left(\frac{32G}{5c^5}\mu_{12}^2 n_1 n_2 v_{12}^5 \sum \frac{d\sigma}{d\Omega} \sin^2\theta\right) V\nu \qquad \ldots(5)$$

The number density will be increased by a factor of gamma and the volume associated with the GRB, will be given by the deceleration volume.

In gamma ray bursts, due to the general relativistic effects, the shock wave propagated from the burst will be decelerated. The blast wave will form a spherical shell around the blast. The radius of this shell is called the deceleration radius.

To determine this consider the energy of GRB given by [7],

$$E_\nu = \frac{4}{3}\pi R_D^3 \Gamma^2 n m_P c^2 \qquad \ldots(6)$$

Hence the volume is given by:

$$V = \frac{E_\nu}{\Gamma^2 n m_P c^2} \qquad \ldots(7)$$

For $n = 10^8 / m^3, E_\nu \approx 10^{44} J$, the volume is of the order of $V = \frac{5 \times 10^{46}}{\Gamma^2} m^3$.

This gives the power of the thermal gravitational waves emitted from a GRB as:

$$\dot{E} = \left(\frac{32G}{5c^5}(\Gamma\mu_{12})^2 \left(\frac{N}{V}\right)^2 v_{12}^5 \sum \frac{d\sigma}{d\Omega}\sin^2\theta\right) V\nu \qquad \ldots(8)$$

Putting in the value for the volume and $\frac{d\sigma}{d\Omega}$, we get:

$$\dot{E} = \left(\frac{32G}{5c^5}(\Gamma\mu_{12})^2 \left(\frac{N}{\left(\frac{E_\nu}{\Gamma^2 n m_P c^2}\right)}\right)^2 v_{12}^5 \left(\frac{e^4}{(8\pi\varepsilon_0)^2 \mu_{12}^2 v_{12}^4}\right)\right)\left(\frac{E_\nu}{\Gamma^2 n m_P c^2}\right)\nu \qquad \ldots(9)$$

For $n = 10^8 / m^3, E_\nu \approx 10^{44} J$, we get the power as, $\dot{E} \approx \Gamma^4 (3.5 \times 10^8) Watts$.

A gamma ray burst corresponding to $\Gamma = 100$, the power works out to be of the order of $\dot{E} \approx 3.5 \times 10^{16} W$ at the frequency of $\nu = \Gamma\left(\frac{kT_c}{h}\right) \approx 10^{22} Hz$.

### **4. Thermal gravitational waves from short duration GRB**

Short gamma ray bursts have a shorter duration (<0.2 – 2 s) and a harder spectrum as compared to the duration of 2 – 200 s for long GRBs. The first of these short GRBs (GRB050509b) was identified with the halo of an elliptical galaxy at a distance of 1.12 Gpc. [7, 8]

Short GRBs are due to the merger of two neutron stars, where as, the long GRBs are due to the collapse of very massive stars. The spectrum observed is harder because the objects merging to produce the GRB are more compact.

The time taken for the merging of two NS is given by,

$$t_{merge} = \frac{GM^2/R}{\dot{E}} \qquad \ldots(10)$$

Where, $\dot{E}$ is the loss of energy due to the emission of thermal gravitational waves.

For two neutron stars of mass $1.5 M_\Theta$ the merger time typically works out to be of the order of $10^9$ years. Due to the longer merger time of the NS, the short GRBs are found in older population elliptical galaxies. [8]

In the case of the merger of two neutron stars, the number density as well as the temperature is substantially high compared to the long duration GRB.
The temperature is of the order of $10^{13} K$ and $n \approx 10^{40}/cc$.
Considering all the gamma factors associated with the GRB [9, 21], as in the previous case, the gravitational power is given by equation (9). Putting in the numbers corresponding to short GRB we get:

$$\dot{E} \approx \Gamma^4 (4 \times 10^{36}) Watts \qquad \ldots(11)$$

At the frequency of $\nu = \Gamma \left( \frac{kT_c}{h} \right) \approx \Gamma(10^{23}) Hz$

During the short duration burst, the two neutron stars undergo collision. During the tidal break up of the neutron stars, its binding energy is released.

For a neutron star of the mass of $1.5 M_\Theta$ and radius of about 10 km, the binding energy released is of the order of:

$$B.E = 2 \left( \frac{3}{5} \frac{GM_{NS}^2}{R_{NS}} \right) \approx 6 \times 10^{46} J \qquad \ldots(12)$$

For a gamma factor of about 100, the power radiated due to the thermal gravitational wave emission is of the order of,

$$\dot{E} \approx 4 \times 10^{44} Watts \qquad \ldots(13)$$

This implies that about 1% of the energy released in the short duration gamma ray burst could be in the form of thermal gravitational waves.

For a typical distance of 100 $Mpc$ for the GRB, the flux is given by:

$$f = \frac{E}{4\pi d^2} \approx 4 \times 10^{-6} W/m^2 \qquad \ldots(14)$$

Since the event occurs over a time scale of one second, the flux is equivalent to the fluence.

Another possible source of thermal gravitational radiation is from Hawking evaporation of primordial blackhole (PBH) of small mass. For a typical PBH, mass $\approx 10^{12} kg$, the Hawking temperature would be $\approx 10^{12} K$, which would imply a thermal flux of $\approx 10^6 W$, which over its lifetime would be $\approx 10^{23} W$. However as lifetime of such PBH's scale as $M^3$, for small PBH's the total energy would be much smaller.

## 5. Thermal gravitational waves from jets

The effect of spin of the black hole (Kerr back holes) produces a very interesting phenomenon of powerful bi-directional jets. The process involves a supermassive black hole that is being fed with magnetized gas through an orbiting accretion disk. The combination of the strong gravity of the black hole, the rotation in the in falling matter, and the magnetic field are responsible for the jet creation. [9]

The jet is powered by both the energy of accretion in the disk and from the rotational energy of the black hole. It is the rotating black hole; however, that provides most of the energy.

The length of the jet in terms of the mass of the central black hole and the number density of the ambient gas is given by, $l = \left( \dfrac{3GM^2}{\pi \rho c^2 (\tan 5)^2} \right)^{1/4}$ \qquad \ldots(15)

The corresponding volume of the jet in terms of the mass of the central black hole and the number density of the ambient gas is:

$$V = \frac{1}{\Gamma}\left(\frac{M}{\sqrt{n}}\right)^{3/2} \qquad \ldots(16)$$

The power associated with the thermal gravitational waves from the jet is given by:

$$\dot{E} = \left(\frac{32G}{5c^5}(\Gamma\mu_{12})^2 \left(\frac{N}{\frac{1}{\Gamma}\left(\frac{M}{\sqrt{n}}\right)^{3/2}}\right)^2 v_{12}^5 \left(\frac{e^4}{(8\pi\varepsilon_0)^2 \mu_{12}^2 v_{12}^4}\right)\right) \left(\frac{1}{\Gamma}\left(\frac{M}{\sqrt{n}}\right)^{3/2}\right) v \qquad \ldots(17)$$

For a given black hole, the length of the jet depends on the density of the particles emitted out along the jet. From our earlier discussion we have,

$$l = \left(\frac{3GM^2}{\pi c^2 (\tan 5)^2}\right)^{1/4} \frac{1}{\rho^{1/4}} \qquad \ldots(18)$$

For a 30 solar mass black hole, with number density of $n = 10^3 \, m^{-3}$, the length of the jet is of the order of, $l \approx 1 kpc$. For a billion solar mass black holes, the variation of the length of the jet with respect to the number density of particles in the jet.

Putting in the values we get: $\dot{E} \approx \Gamma^3 (6.5 \times 10^{20}) J/s$ …(19)

Here we notice $\Gamma^3$ dependence for the thermal gravitational waves from a jet, where as we had $\Gamma^4$ dependence in the case of gamma ray bursts.

### 6. Detection of thermal gravitational waves

When a weak gravitational wave passes through a gas along a line perpendicular to the plane of the particles then the particles will oscillate. The area enclosed by the particles does not change, and there is no motion along the direction of propagation. Passing of gravitational waves through a system of mass sets it into harmonic oscillations, hence causing a strain $h$ of the order of $\approx 10^{-22}$. [10]

However, current devices in operation like LIGO are expected to detect an $h \approx 10^{-22}$. [11] But what about the detection of thermal gravitational waves. So far there have been few attempts to conceive detection of such waves. We have estimated earlier that the flux of thermal gravitational wave from the sun (around frequency of $\approx 10^{16} Hz$) at earth is about 0.5 watts. How can we detect this kind of high frequency gravitational wave radiation? For example, in magnetised plasma, gravitational waves can be coupled to electromagnetic waves and can get damped. [12]

The damping time is given by: $\tau = \dfrac{\omega_B}{GnT^{1/2}m_n^{3/2}}$ …(20)

For a magnetic field of $10^{15} G$ and $T \approx 10^{12} K$, the damping time is of the order of $10^2 \sec$.

If two particles in the gas is separated by a distance of *d*, and the passing of gravitational wave causes a strain *h*, then the change in the distance between the particles is given by:

$\Delta d = hd$

The distance of separation is given by, $d = \dfrac{v}{\omega}$ where, $\omega$ is the frequency and $v = \sqrt{\dfrac{kT}{m}}$ is the velocity. The change in velocity due to the passing of the gravitational wave is given by:

$\Delta v = hd\omega$ …(21)

The energy change per collision is given by:

$dE = \dfrac{1}{2}m\left(\dfrac{\Delta v}{v}\right)^2$ …(22)

If the number density of the gas is *n* and $\lambda$ is the mean free path, the total power radiated by the volume is given as:

$\dot{E} = \dfrac{1}{2}m\left(\dfrac{\Delta v}{v}\right)^2 \lambda n = \dfrac{m^2 n\lambda}{2}\dfrac{(hd\omega)^2}{kT}$ …(23)

The energy density of the wave is of the order of: $\dfrac{h^2\omega^2 c^2}{G}$ …(24)

On integrating the expression for power we get the time scale for the damping of the wave as: [12]

$$\tau = \lambda \left(\frac{c}{v}\right)^3 \left(\frac{\omega^2}{Gmn}\right)$$

$$\tau = \lambda \left(\frac{c}{v}\right)^3 \left(\frac{\omega^2}{\omega_G^2}\right) \qquad \ldots(25)$$

The quantity $(Gmn)$ has the same dimensions as $\omega^2$, hence $\sqrt{Gmn}$ can be interpreted as the gravitational plasma frequency $\omega_G^2$ associated with the gas undergoing oscillations due to the passing of the gravitational waves.

In the case of neutron stars with temperature of the order of $T \approx 10^{11} K$ we have $\omega \approx \omega_G$.

The damping time will be of the order of $10^{-18}$ sec and hence waves may be trapped within the star. The trajectories of charged particles may be affected by passage of gravitational waves, which involves generation of electric current in the magnetised plasma.

The high frequency gravitational waves can also be detected through the atomic transitions induced by them at a very slow rate. [1]

The quadrupole transition of hydrogen from $3d \rightarrow 1s$ state with emission of a graviton occurs at a frequency of $\approx 10^{15} Hz$.

The quantum mechanical transition rate is given by: [13] $\Gamma = \dfrac{P}{\hbar\omega}$ ...(26)

Where, $P$ is the power emitted by a dipole. For the case of spontaneous graviton emission, the quadrupole gravitational power is given by:

$$P = \frac{2G\omega^6}{5c^5} I^2 \qquad \ldots(27)$$

The normalised wavefunctions for the 1s and 3d states is given by:

$$\psi_{1s} = \frac{1}{\sqrt{\pi}a^{3/2}}\exp(-r/a); \quad \psi_{3d} = \frac{1}{162\sqrt{\pi}a^{3/2}}\left(\frac{r^2}{a^2}\right)\exp(-r/3a)\sin^2\theta \qquad \ldots(28)$$

Where, $a = \frac{\hbar}{m_e e^2}$ is the Bohr radius.

Using these expressions in transition rate, we get: [13]

$$\Gamma = \frac{P}{\hbar\omega} = \frac{\alpha^2 G m_e^3 c}{360\hbar^2} \approx 6\times 10^{-40} \text{ sec}^{-1} \qquad \ldots(29)$$

And the corresponding lifetime of the transition is:

$$\tau = \frac{5\hbar^2}{256 G c \alpha^6 m_e^3} \approx 10^{36} \text{ sec} \qquad \ldots(30)$$

The frequency of this transition is of the order of $10^{16}$ Hz, which is within the range of thermal gravitons emitted from the sun. About $10^3$ gravitons $m^{-2}s^{-1}$ fall on the earth from the sun at this frequency. Thus there is a finite probability of detecting induced emission with a sufficiently large detector. This radiation will be very penetrating.

By coincidence, the lifetime for this transition to take place is the same as the proton decay time ($\approx 10^{31}$ yr ~ $10^{38}$ s). Proton decay is a major prediction of GUTs and despite the long lifetime is being tested by several experiments. So it may not altogether be impossible to also observe the effects of high frequency thermal gravitational radiation, which can also induce transitions with a lifetime comparable to that of proton decay. [1]

The absorption rate of these high frequency gravitons can be estimated as $\approx 10^{-27}$, for terrestrial detectors, so one requires a detector of several hundred square kilometres to detect a few transitions in some decades. [20]

Another way of detecting these thermal radiations is to convert them into electromagnetic waves of same frequency. When an electromagnetic wave of amplitude $H_Y$ propagates through a constant magnetic field $H_0$, produces a quadrupole stress term given by:

$$T_{YY} = H_Y H_0 \cos(kx - \omega t) \quad \ldots(31)$$

This stress term gives rise to gravitational waves given by the linear Einstein equation: [14, 15]

$$\Box h_{YY} = kT_{YY}; \quad k = \frac{16\pi G}{c^4} \quad \ldots(32)$$

Alternatively, a weak gravitational wave $h_{YY}$ propagating through a magnetic field $H_0$ gives rise to a magnetic field perturbation given by:

$$\Box H_Y = \omega^2 h_{YY} H_0 \quad \ldots(33)$$

The fraction of the gravitational wave energy converted into electromagnetic waves of frequency $\omega$ is given by: [16, 17] $f = kH_0^2 d^2$ ...(34)

Where, d is the special extent of the uniform magnetic field.

## 7. Background thermal gravitational radiation

At the Planck epoch of $t \approx 10^{-43} s$, all the interactions were of equal strength so that thermal equilibrium was maintained between gravitons and other particles. As the universe expanded, the gravitational interaction weakened and graviton decoupled from other particles. [2]

If N is the number of particles that were coupled with the graviton, then the temperature of the background gravitational radiation is given by:

$$T_g = \left(\frac{43}{22N}\right)^{1/3} T_r \quad \ldots(35)$$

Where, $T_r$ is the background radiation temperature. And for the present radiation temperature of about 2.7 Kelvin and $N \approx 30$, we have the temperature of the background gravitational radiation of 1 Kelvin.

The implication of this result is in the confirmation of the inflationary model of the universe, according to which the universe went through an exponential expansion of $R \sim \exp(\sqrt{\Lambda} t)$.

At the time of inflation, the expansion of the universe occurred at an exponential rate, where the expansion was by a factor of $10^{28}$. The expansion time of the universe when the inflation occurred is about $10^{-36} s$. [18]

The radiation temperature at this epoch is given by:

$$T = \left(\frac{3c^2}{32\pi Ga}\right)\frac{1}{t^{1/2}} \approx \frac{2\times 10^{10}}{t^{1/2}} \approx 2\times 10^{28} K \qquad \ldots(36)$$

The temperature of the background thermal gravitational radiation corresponding to this radiation temperature is given by:

$$T_g = \left(\frac{43}{22N}\right)^{1/3} T_r \approx 8\times 10^{27} K \qquad \ldots(37)$$

And the corresponding wavelength is:

$$\lambda = \frac{hc}{kT_g} \approx 10^{-30} m \qquad \ldots(38)$$

At the end of the inflation phase, the wavelength of the gravitational radiation background is $10^{-2} m$. However, in order not to interfere with nucleosynthesis in the hot dense phase, its energy density would have to be $<1\%$ of the radiation energy density. This would give an $h$ of the order of $\frac{h^2\omega^2 c^2}{G}$.

Now the wave would be stretched by a further factor of $10^{19}$. This would give a wavelength of $10^{17} m$ at present. Detection of such waves through fluctuations in the cosmic microwave background radiation could verify the existence of such a phase in the early universe. [19]

If inflation had not taken place, we would be left with a thermal gravitational wave background with a temperature of $\sim 1K$. If at all this can be detected, it would provide evidence against inflation. [1, 2]

Also the detection of these thermal background gravitational waves provides a basis to verify the validity of the big bang model itself. There are claims that the most convincing evidences of the big bang, the microwave background and abundance of helium, can be accounted for without invoking the big bang.

But these thermal gravitational waves cannot be generated without the universe passing through the super hot, super dense Planck epoch.

Interestingly enough, equation (34) can be used to put a limit on any primordial intergalactic magnetic field. For instance, if $H_0 = 0.1G$ at recombination ($z \sim 10^3$), then $f \sim 10^{-3}$.

Larger values of $f$ would cause anisotropies in the photon background larger than measured and would result in a noticeable weakening of a single polarization of the microwaves. This already limits present day $H_0$ to less than $10^{-6}G$ in intergalactic space, or less than $10^{-4}G$ in interstellar space. The number density of the graviton background is of the order of $10^9/m^3$ so that their flow through a square kilometre area is of the order of $10^{24} s^{-1}$. [1, 18]

As we have seen in section [2], the power of thermal gravitational waves emitted by a neutron star is of the order of $10^{22}W$. A neutron star near the galactic centre (about a kilo parsec from earth) would produce a flux of approximately same order as that of the sun. If the neutron star is emitting these thermal gravitational waves from a nearby source, then the flux could be as high as $100W$.

From equation (34) we can determine the fraction of the energy converted to electromagnetic radiation. For a neutron star of magnetic field $10^9 G$ and of size of about 10 kilometres, the fraction is given by:

$f = kH_0^2 d^2 \approx 10^{-16}$, Where, $k = \dfrac{16\pi G}{c^4}$.

Hence for the neutron star about $10^{15} J$ of energy will be converted to high-energy gamma rays.

## 8. Integrated background thermal gravitational radiation (IBTGR)

Thermal gravitational waves continue to be generated in the early universe as it expands, as the temperature and particle densities continue to be high. We can estimate the integrated power emitted in thermal gravitational waves as the universe cools from say $10^{13} K - 10^5 K$. [2, 10, 14]

This is similar to what was done for stellar core but now all the quantities are time dependant. The time range, corresponding to the above temperatures is $10^{-6} - 10^{10} s$.

The power per unit volume per unit frequency interval is given by the quadrupole formula as given in equation (1):

$$\dot{E} = \left( \frac{32G}{5c^5} \mu_{12}^2 n_1 n_2 v_{12}^5 \sum \frac{d\sigma}{d\Omega} \sin^2 \theta \right) V \nu$$

The differential cross section is given by equation (2) as:

$$\sum \frac{d\sigma}{d\Omega} \sin^2 \theta = \frac{e^4}{(8\pi\varepsilon_0)^2 \mu_{12}^2 v_{12}^4}$$

The number density is dependent on time as:

$$n \propto t^{-2} \Rightarrow n = n_i \left( \frac{t_i}{t} \right)^2 \qquad \ldots(39)$$

Where the quantities with suffix '$i$' indicates the initial values of theses quantities at time $10^{-6} s$.

The frequency and the velocity is given by: $\nu = \frac{kT}{h}$ and $v_{12} = \sqrt{\frac{3kT}{\mu_{12}}}$.

The temperature dependence on time is given by: $T = \dfrac{10^{10}}{t^{1/2}}$ ...(40)

Hence the time dependence of frequency and velocity is given as:

$$\nu = \nu_i \left(\dfrac{t_i}{t}\right)^{1/2} \quad ...(41)$$

$$v = v_i \left(\dfrac{t_i}{t}\right)^{1/4} \quad ...(42)$$

The volume of the universe is given by: $V = 2\pi^2 R^3$.

The radius of the universe is related to the temperature as: $RT = $ constant.

Hence we have the volume given by:

$$V = V_i \left(\dfrac{T_i}{T}\right)^3 \Rightarrow V = V_i \left(\dfrac{t}{t_i}\right)^{3/2} \quad ...(43)$$

Using equations (39) – (43) in the expression for power radiated by the thermal background gravitational radiation we have:

$$\dot{E} = \left(\dfrac{32G}{5c^5}\right) n_i^2 v_i \dfrac{e^4}{(8\pi\varepsilon_0)^2} V_i \nu_i \left(\dfrac{t_i}{t}\right)^{3/4} \quad ...(44)$$

The initial values corresponding to time $t = 10^{-6} s$ is given by:

$$n_i = n\left(\dfrac{t}{t_i}\right)^2 = 1\, proton/m^3 \left(\dfrac{13.7 \times 10^9 \times 3.15 \times 10^7}{10^{-6}}\right)^2 \quad ...(45)$$

$$\nu_i = \dfrac{kT_i}{h} \approx 2 \times 10^{23}\, Hz \quad ...(46)$$

$$v_i = \sqrt{\dfrac{3kT_i}{m_P}} \approx 3 \times 10^8\, m/s \quad ...(47)$$

$$R_i = R\left(\dfrac{T}{T_i}\right) = \left(3 \times 10^8 \times 13.7 \times 10^9 \times 3.15 \times 10^7\right)\left(\dfrac{2.7}{10^{13}}\right)$$

$$V_i = 2\pi^2 R_i^3 \approx 10^{42}\, m^3 \quad ...(48)$$

Using equations (45) – (48) in equation (44) we get:

$$\dot{E} = \frac{2 \times 10^{56}}{t^{3/4}} \qquad \ldots(49)$$

The integrated energy is given by:

$$\int \dot{E} dt = \int_{10^{-6}}^{10^{10}} \left( \frac{2 \times 10^{56}}{t^{3/4}} \right) dt \approx 10^{57} t^{1/4} \Big|_{10^{-6}}^{10^{10}} \approx 10^{60} J \qquad \ldots(50)$$

This is the integrated energy radiated by the background thermal gravitational radiation in the early universe as the temperature cooled from $10^{13} K - 10^{5} K$.

As the universe expanded, this energy is red shifted by a factor given by:

$$\frac{R(present)}{R(T = 10^5 K)} = \frac{10^5}{2.7} \approx 10^5 \qquad \ldots(51)$$

Therefore the present energy associated with the IBTGR is of the order of $10^{55} J$.

If $\varepsilon = \frac{E}{V_P}$ is the energy density, where, $V_P = 2\pi^2 R_P^3 \approx 4 \times 10^{79} m^3$ is the present volume of the universe, then the flux associated with this IBTGR is given by:

$$f = \varepsilon \frac{c}{4} \approx 2 \times 10^{-17} J/m^2/s \qquad \ldots(52)$$

The energy density associated with the cosmic microwave background radiation is of the order of $10^{-14} J/m^3$. The corresponding flux is given by:

$$f_{CMB} = 10^{-14} \times \frac{3 \times 10^8}{4} \approx 10^{-6} J/m^2/s \qquad \ldots(53)$$

From equations (52) and (53) we see that the flux associated with IBTGR is about 10 orders less than that of the CMB radiation.

The contribution of the microwave background radiation to the normalised critical density of the universe is $\Omega_{CMB} \approx 4 \times 10^{-5}$. Since the flux associated with IBTGR is about 10 orders less the contribution to $\Omega$ due to this will be $\Omega_{IBTGR} \approx 10^{-15}$.

As we have seen in section [1], the power of thermal gravitational waves emitted by a star is of the order of $10^9 W$. Considering all the $10^{11}$ stars in the $10^{11}$ galaxies, the power associated with the thermal gravitational waves from all these stars is of the order of:

$$\dot{E} = 10^{22} \times 10^9 = 10^{31} W \qquad \ldots(54)$$

The total energy emitted due to the thermal gravitational waves by these stars over their average life span of $\approx 3 \times 10^{17} s$ is given by:

$$E \approx 10^{31} \times 3 \times 10^{17} \approx 3 \times 10^{48} J \qquad \ldots(55)$$

The power of thermal gravitational waves emitted by neutron stars (as discussed in section [2]) is of the order of $10^{22} W$.

If one out of every 100 stars is a neutron star, then the power associated with the thermal gravitational waves from all the neutron stars is:

$$\dot{E} = 10^9 \times 10^{11} \times 10^{22} = 10^{42} W \qquad \ldots(56)$$

The total energy emitted due to the thermal gravitational waves by these neutron stars over their average life span of $\approx 10^8 s$ is given by:

$$E \approx 10^{42} \times 10^8 \approx 10^{50} J \qquad \ldots(57)$$

As for the PBH's, we need something like $10^{30}$ of them to match the thermal gravitational wave background. This would need a density of such objects much more than what is implied by the gamma ray background.

The overall energy associated with the emission of thermal gravitational waves from stellar sources is of the order of $10^{50} J$, which is still about 5 orders less than that associated with IBTGR as we have seen earlier in the section.

## 9. Concluding remarks

We have given detailed estimate of the high frequency thermal gravitational radiation flux from a variety of high energy sources including nascent neutron stars, gamma ray bursts, relativistic jets, evaporating primordial black holes, etc. we have also estimated the integrated thermal gravitational wave flux in the expanding early universe. This integrated thermal gravitational wave background has a present energy density about five orders of magnitude higher than that from all other discrete sources.